\documentclass[%
 reprint,
 amsmath,amssymb,
 aps,
]{revtex4-2}

\usepackage{graphicx}
\usepackage{dcolumn}
\usepackage{bm}
\usepackage{lineno}
\usepackage{adjustbox}
\usepackage{epsfig}
\usepackage{epstopdf}
\usepackage{float}
\usepackage{amssymb}
\usepackage{color}
\usepackage{bm}
\usepackage{amsfonts}
\usepackage{lineno,hyperref}
\usepackage{array}
\usepackage{microtype}
\usepackage{xcolor}
\usepackage{caption}
\usepackage{subcaption}

\begin{document}

\preprint{APS/123-QED}

\title{Dynamical stability and phase space analysis of an Emergent Universe with non-interacting and interacting fluids}
\author{Bikash Chandra Roy $^1$}%
\email{bcroy.bcr25@gmail.com}

\author{Anirban Chanda $^1$}
\email{aniphys93@nbu.ac.in}

\author{Bikash Chandra Paul $^1$}
\email{bcpaul@nbu.ac.in}
\affiliation{$^1$Department of Physics, University of North Bengal, Raja Rammohunpur, 734013, India}

%

\date{19/11/2023}

\begin{abstract}
We investigate the evolution of a flat Emergent Universe obtained with a non-linear equation of state (nEoS) in Einstein's general theory of Relativity. The nEoS is equivalent to three different types of barotropic cosmic fluids, which are found from the nEoS parameter. The EU began expanding initially with no interaction among the cosmic fluids. Assuming an interaction that sets in at a time $t \geq t_i$ in the fluid components, we study the evolution of the EU that leads to the present observed universe. We adopt a dynamical system analysis method to obtain the critical points of the autonomous system for studying the evolution of an EU with or without interaction in fluid components.  We also study the stability of critical points and draw the phase portraits. The density parameters and the corresponding cosmological parameters are obtained for both the non-interacting and interacting phases of the evolution dynamics.
\end{abstract}

\maketitle

\section{Introduction}
\label{sec:intro}

Cosmology has transitioned from speculative science to experimental science with the advent of precision measurements from different cosmological missions over the last few decades. The present-day cosmological observations made it clear that the observed universe is not only expanding but is accelerating \cite{riess_observational_1998,1999ApJ...517..565P, Bennett_2003,riess2004type,eisenstein_detection_2005,bennett_nine-year_2013}. It is known that the most successful theory to describe gravitational interaction is Einstein's theory of General Relativity (GR) and based on GR, the Lambda Cold Dark Matter ($\Lambda$CDM) model is the most acceptable cosmological model. It accommodates the accelerating expansion of the universe and matches well with astronomical observations. Although the standard model agrees with most of the recent cosmological observations, it is plagued with several problems, namely, the horizon problem, flatness problem, singularity problem, etc. \cite{brandenberger_inflationary_1999,kolb_early_2019}. To resolve several of these early universe issues, the idea of cosmic inflation is proposed in the literature, where a rapid expansion of space in the early universe engulfed the entire space in the universe in a very short interval of time. Guth, Linde and others show that a homogeneous scalar field in the framework of standard cosmology permits such inflation \cite{Guth1981,sato_first-order_1981,linde2002inflation}. Furthermore, it was shown that inflation can describe the large-scale structure formation of the universe. It is known that the $\Lambda$CDM model also suffers from some conceptual issues, namely, the cosmological constant problem, the coincidence problem, the Hubble tension, the $\sigma_{8}$ tension, etc. \cite{sahni_case_2000,carroll_cosmological_2001,padmanabhan_cosmological_2003,peebles_cosmological_2003,freedman_carnegie-chicago_2019,yuan_consistent_2019,freedman_measurements_2021,riess_cosmic_2021,soltis_parallax_2021}. Consequently a number of papers \cite{sotiriou_fr_2010, De_Felice_2010, Nojiri_2011, Nojiri_2017, PhysRevD.84.024020, NOJIRI20051,ferraro_modified_2007,ferraro_born-infeld_2008, bengochea_dark_2009, Jim_nez_2018, Maartens_2010, wang_horava_2017} came up to resolve the above issues with a modified theories of gravity and several GR-based models beyond the $\Lambda$CDM model \cite{Kamenshchik_2001, Bento_2002, Benaoum_2002, chiba_quintessence_1999, amendola_coupled_2000, martin_quintessence_2008, Nojiri_2004, Nojiri_2005_1, COPELAND_2006, Capozziello_2006, Capozziello_2006_1, Durrer_2007, bamba2012dark, Odintsov_2017, Nojiri_2021}, that aim to resolve the above issues.

Later modifications beyond $\Lambda$CDM lead to different dynamical dark energy (DE) models, namely, Chaplygin gas \cite{Kamenshchik_2001} and its variations \cite{Bento_2002}, \cite{Benaoum_2002}, models consisting of one or more scalar fields namely, quintessence \cite{chiba_quintessence_1999}, \cite{amendola_coupled_2000}, \cite{martin_quintessence_2008}, etc. are considered. A detailed review of different DE models having fluids with non-linear equations of state which includes quintessence, K-essence, Tachyon, Phantom, etc., can be found in the Refs. \cite{Nojiri_2004, Nojiri_2005_1, COPELAND_2006, Capozziello_2006, Capozziello_2006_1, Durrer_2007,  bamba2012dark,  Odintsov_2017, Nojiri_2021}. The modified theories of gravity were employed to explore the connection between modification of the early inflationary phase with the late time acceleration phase \cite{nojiri_modified_2006}. In the literature, different modified gravitational theories are considered to explain several astrophysical and cosmological phenomena and the viabilities of these models are also tested using astronomical observations \cite{starobinsky198030}, \cite{RUDRA2021115428}, \cite{PhysRevD.76.044027}, \cite{mandal2020cosmography}.

In the standard model, the existence of the initial Big Bang singularity is a crucial issue in constructing a cosmological model since it can be probed to know whether our universe originated from a singularity at the beginning or has always existed. The lack of a solution to this fundamental issue in the setting of inflationary cosmology motivated several authors to develop pre-inflationary scenarios or initially an emergent universe (EU)  with no singularity\cite{Ellis2003, Ellis_2003} and cyclic scenarios\cite{PhysRevD.65.126003}, which are typically non-singular, a signature of the past eternal. Some alternative singularity-free cosmological models are found in the literature inspired by the generally accepted theory that incorporating quantum gravity (QG) phenomena at a very low scale leads to the natural disappearance of singularities \cite{Bojowald_2001, Hossain_2010}.

The  EU scenario, which originated from a non-singular state, is one of the prominent possibilities that has been given careful consideration by various authors \cite{Ellis2003, Ellis_2003,mukherjee2005emergent,Mukherjee_2006,paul_emergent_2021}. It is known that the EU scenario was first proposed by Ellis and Maartens to avoid the singularity problem of Big Bang cosmology \cite{Ellis2003}. In an EU, the universe emerges as an Einstein static universe in the infinite past ($t \rightarrow - \infty$) and avoids the initial singularity by staying large at all times. The universe gradually expands slowly to attain a Big Bang phase of expansion. In the EU model, an inflationary universe emerges from a static phase and eventually leads to a macroscopic universe that occupies the present observed universe.  Once inflation starts, it remains in that phase, explaining the present acceleration. Ellis {\it et al.} \cite{Ellis_2003} obtained an EU scenario for a closed ($k=1$, $k$ being the curvature parameter) universe considering a minimally coupled scalar field ($\phi)$ with a special choice of potential where the universe exits from its inflationary phase and gradually reheats as the scalar field starts oscillating around the minimum of the potential. Later, it was shown that such a potential occurs naturally by the conformal transformation of the Einstein-Hilbert action with $\alpha R^2$ term, where $\alpha$ is a coupling constant. Present observations indicate that the universe is almost flat ($k=0$) with a negligible spatial curvature. EU scenario in a flat universe can be obtained in a semi-classical theory of gravity. In the Starobinsky model, Mukherjee {\it et al.} obtained an EU with all its features in a flat Robertson–Walker (RW) spacetime geometry \cite{mukherjee2005emergent}. Mukherjee {\it et al.} \cite{Mukherjee_2006} proposed another interesting class of EU model considering a non-linear equation of state (nEoS) in a flat universe. In this framework, the cosmic fluid is equivalent to a mixture of three different types of fluid, described by the nEoS given by: 
\begin{equation}\label{EQ:11}
p = A\rho - B\sqrt{\rho},    
\end{equation}
where $A$ and $B$ are constant parameters. The composition of the cosmic fluid is determined for a given value of the EoS parameter $A$. Various theories of gravity, such as the Brans–Dicke theory \cite{del2007emergent}, brane world cosmology \cite{Banerjee2008}, Gauss-Bonnet modified gravity \cite{paul2010emergent}, Loop quantum cosmology \cite{mulryne2005emergent}, Energy-momentum squared gravity \cite{khodadi2022emergent}, $f(R,T)$ gravity \cite{debnath_observational_2020}, etc. are among the theories of gravity where EU models are explored. A non-linear sigma model was used to study the EU \cite{beesham_emergent_2009-1}. An EU model with particle creation and irreversible matter creation is also studied by Ghosh and Gangopadhyay using a thermodynamical approach \cite{Ghosh_2017}. The validity of EU models is studied using recent cosmological observations with the estimation of the observational constraints on the model parameters \cite{paul_constraints_2010-1, paul_emergent_2011, ghose_observational_2012-1}. Recently \cite{paul_observational_2019} studied the EU scenario described by a nEoS in addition to viscosity. The observational bounds of the model parameters are determined.

The objective of the paper is to study the EU with or without interacting fluids and to analyze the dynamical systems obtained from the field equations. 
In a non-interacting case for the EU, the cosmic fluids are fixed for a given EoS parameter $A$.
However, for a fluid where interaction sets in after time $t \geq t_i$, the interaction strength plays a crucial role in the evolution of the late universe. The above idea incorporated in the EU model is realistic because there are cosmological models where the cosmic fluid components interact with one another via energy exchange from one sector of the fluid to the other. Recently, different cosmological models with interaction among the dark sectors gained popularity \cite{PhysRevD.73.103520, PhysRevD.81.043525, PhysRevD.81.023007, PhysRevD.83.023528, PhysRevD.85.107302}. It is interesting to note that although the individual fluid components violate the energy conservation equation in the case of interacting cosmology, the total energy density remains conserved. In this case, we consider that the universe evolved from a radiation-dominated epoch to begin with enter into a matter and DE-dominated era when the interaction sets in at late times. As the field equations are highly complex here we adopt the dynamical system analysis technique \cite{boehmer2010jacobi} to investigate the behaviour of the cosmological dynamics of the EU in the presence of non-interaction or interaction. We use the dynamical system analysis to study the behaviour of a cosmological model under small perturbations, which can provide some crucial insights
into the model's viability. Besides this, it contributes to understanding the structure as well as the evolution of the universe. The implications of dynamical system analysis in a cosmological model are studied in the literature \cite{B_hmer_2016, Bhanja_2023}. In the literature, dynamical system analysis in modified theories of gravity is employed for understanding the features of the universe
\cite{Faraoni_2005, Guo_2013, gonçalves2023cosmological, Khyllep_2023, Narawade_2023}. In the paper, we study both the interacting as well as non-interacting cosmic fluids in a flat EU, considering a system of autonomous differential equations. 

The paper is organized as follows: In sec. (\ref{sec:FE}), the basic field equations for the EU are obtained in a homogeneous, isotropic and spatially flat space-time. The energy density and pressure are determined for non-interacting and interacting cases differently in subsections (\ref{non-int}) and (\ref{int}), respectively. Assuming an epoch when interaction sets in say $t>t_{i}$, the cosmic fluids and the conservation equations for the fluid's components are rewritten. The effective EoS parameters in the presence of interaction are determined by the strength of the interaction. In sec. (\ref{sec:dynamic}), The field equations are rewritten as differential equations and the dynamical system analysis methodology is adopted for the study of the autonomous system with the interacting or non-interacting fluids. The evolutionary behaviour of the EU is presented. Cosmological implications of the critical points are also discussed. Finally, we summarize the result obtained here in sec. (\ref{sec:conclusions}).
\section{Field Equations for Emergent universe}
\label{sec:FE}

The Einstein field equation (EFE) is 
\begin{equation}\label{EQ:21}
    R_{\mu\nu} - \frac{1}{2}g_{\mu\nu}R =  T_{\mu\nu},
\end{equation}
where, $R_{\mu\nu}$ is the Ricci tensor, $R$ is the Ricci scalar, $g_{\mu\nu}$ is the metric tensor and $T_{\mu\nu}$ is the energy-momentum tensor of the cosmic fluid. We work in natural units, $c^2= 8\pi G = 1$.

We assume a homogeneous, isotropic and spatially flat spacetime described by the Robertson-Walker (RW) metric, which is 
\begin{equation}\label{EQ:22}
    ds^{2} = - dt^{2} + a^{2}(t)\Big[dr^{2} + r^{2}(d\theta^{2} + \sin^{2}\theta\; d\phi^{2})\Big],
\end{equation}
where $a(t)$ is the scale factor of the universe and $r$, $\theta$, and $\phi$ are the comoving coordinates. 

Using the  metric given by Eq. (\ref{EQ:22}) in EFE and the Energy momentum tensor $T^{\mu}_{\nu} = (\rho, - p, - p, - p) $ where $\rho$ denotes the energy density and $p$ denotes the pressure of the cosmic fluid we get
\begin{equation}\label{EQ:23}
    3H^{2} =  \rho,
\end{equation}
\begin{equation}\label{EQ:24}
    2\dot{H} + 3H^{2} = - p ,
\end{equation}
here, $H = \frac{\dot{a}}{a}$ being  the Hubble parameter. The energy conservation equation for the cosmic fluids is given by
\begin{equation}\label{EQ:25}
    \dot{\rho} + 3H(\rho + p) = 0.
\end{equation}

Using Eqs. (\ref{EQ:23}) and (\ref{EQ:24}) and the nEoS given in Eq. (\ref{EQ:11}), one obtains a second-order differential equation for the scale factor,
\begin{equation}\label{EQ:26}
    2\frac{\ddot{a}}{a}+(3A+1)\left(\frac{\dot{a}}{a}\right)^{2}-\sqrt{3}B\frac{\dot{a}}{a}=0.
\end{equation}
On integrating the above equation twice, we obtain the scale factor ($a(t)$) given by, 
\begin{equation}\label{EQ:27}
    a(t) = \Big[\frac{3K(1+A)}{2}\left(K_{1}+\frac{2}{\sqrt{3}B}e^{\frac{\sqrt{3}Bt}{2}}\right)\Big]^{\frac{2}{3(1+A)}},
\end{equation}
where $K$ and $K_{1}$ are the two integration constants. It is evident that it leads to a singular universe if $B<0$, but one gets a nonsingular solution in the case  $B>0$ and $A>-1$. The latter solution is interesting, which leads to an emergent universe (EU) that emerges from an initial Einstein static phase. The scale factor $a(t)$ remains finite even at an infinite past ($t \rightarrow - \infty$).
\subsection{Non-interacting fluids}
\label{non-int}
To begin with, we consider that the cosmic fluid described by the nEoS is not interacting.  Consequently, the conservation equation given by Eq.(\ref{EQ:25}) and Eq. (\ref{EQ:11}) yields  the energy density as follows:
\[
\rho = \frac{B^{2}}{(1+A)^{2}} + \frac{2BK}{(1+A)^{2}}\frac{1}{a^{\frac{3}{2}(1+A)}}
\]
\begin{equation}\label{EQ:211}
 \hspace{3.5 cm}   + \frac{K^{2}}{(1+A)^{2}}\frac{1}{a^{3(1+A)}},
\end{equation}
Therefore we find that the energy density is equivalent to three different fluids given by : ($\rho_{1} = \frac{B^{2}}{(1+A)^{2}}$, $ \rho_{2} = \frac{2BK}{(1+A)^{2}}\frac{1}{a^{\frac{3}{2}(1+A)}}$ and $\rho_{3} = \frac{K^{2}}{(1+A)^{2}}\frac{1}{a^{3(1+A)}}$). Now, using the energy density in Eq. (\ref{EQ:11}) we determine the pressure, which is
\[
p = -\frac{B^{2}}{(1+A)^{2}}+\frac{BK(A-1)}{(1+A)^{2}}\frac{1}{a^{\frac{3}{2}(1+A)}}
\]
\begin{equation}\label{EQ:212}
  \hspace{3.5 cm}   +\frac{AK^{2}}{(1+A)^{2}}\frac{1}{a^{3(1+A)}},
\end{equation}
it contains three different terms that correspond to three different types of fluids, 
namely, $p_{1}=-\frac{B^{2}}{(1+A)^{2}}$, $p_{2}=\frac{BK(A-1)}{(1+A)^{2}}\frac{1}{a^{\frac{3}{2}(1+A)}}$ and $p_{3}=\frac{AK^{2}}{(1+A)^{2}}\frac{1}{a^{3(1+A)}}$ are identified with different barotropic fluids depending on $A$. Comparing with the barotropic EoS given by $p_{i} = \omega_{i}\rho_{i}$ we  get  $\omega_{1} = - 1$, $\omega_{2} = \frac{A-1}{2}$ and $\omega_{3} = A$. The first term can be interpreted as a cosmological constant that accommodates the DE sector of the universe. The parameter $A$ plays an important role in determining the composition of the fluids in the universe. In Table (\ref{tab1}), the composition of the cosmic fluids is shown for different values of $A$ parameter (see also \cite{Mukherjee_2006}). So, for a specific value of $A$, the composition of the cosmic fluid is fixed in the case of non-interaction.

\begin{table*}
\centering
\caption{Composition of universal matter for various values of A}
\begin{tabular}{ |c | c | c | c | c | c |}
\hline
 A & $\frac{\rho_{2}}{\Lambda}$  & $\omega_{2}$ & $\frac{\rho_{3}}{\Lambda}$ & $\omega_{3}$  & Fluid Compositions\\
  & in unit $\frac{K}{B}$ &  & in unit $(\frac{K}{B})^{2}$ &   &  \\ 
\hline
 $\frac{1}{3}$ & $\frac{9}{8a^{2}}$ & $- \frac{1}{3}$ & $\frac{9}{8a^{4}}$ & $\frac{1}{3}$  & dark energy,\\
 &  &  &  &  & cosmic string (CS) and radiation \\
\hline
 -$\frac{1}{3}$ & $\frac{9}{2a}$ & $- \frac{2}{3}$ & $\frac{9}{4a^{4}}$ & - $\frac{1}{3}$  & dark energy,\\
 &  &  &  &  & domain wall (DW) and cosmic string \\
\hline
 $1$ & $\frac{1}{2a^{3}}$ & $0$ & $\frac{1}{4a^{6}}$ & $1$  & dark energy,\\
 &  &  &  &  & dust and stiff matter (SM) \\
\hline
 $0$ & $\frac{2}{8a^{3/2}}$ & $- \frac{1}{2}$ & $\frac{1}{a^{3}}$ & $0$  & dark energy,\\
 &  &  &  &  & exotic matter (EM) and dust \\
\hline
\end{tabular}
\label{tab1}
\end{table*}
Thus, there is a limitation in the case of noninteracting fluids. Once $A$ is fixed, the types of fluids in the universe are fixed. However, in the presence of interacting fluids, the evolution of the universe will be interesting.  When interaction sets in with a given strength of interaction at a later epoch, it is possible to transform a universe with a composition of flids to another which encompass the present universe. 

\subsection{Interacting fluids}
\label{int}
In this subsection, we explore the effect of interaction among the cosmic fluids. It is known that in the early universe, interactions may have originated among the cosmic fluids because of different reasons. The matter-energy content of the universe is fixed throughout the universe's evolution in the non-interacting case of EU for a fixed value of $A$ \cite{Mukherjee_2006}. The composition of cosmic matter changes and different components dominate at different epochs of the universe which is shown in an interacting fluid scenario \cite{Paul_2015} earlier. 

We assume the interaction among the fluids that sets in at $t>t_{i}$, where $t_{i}$ is the time when interaction began. We start with $A = \frac{1}{3}$, which corresponds to a universe with  DE, cosmic string, and radiation with no interaction.  In this subsection, for exploring the dynamical evolution of an EU we assume an interaction that may originated between the DE and cosmic string (CS) sectors and radiation at $t \geq t_i$. 
The energy densities of DE and CS satisfied the following conservation equations \cite{PhysRevD.73.103520, PhysRevD.81.043525, PhysRevD.81.023007, PhysRevD.83.023528},
\begin{equation}\label{EQ:221}
\dot{\rho}_{1} +3H(\rho_{1} + p_{1}) =  Q,   
\end{equation}
\begin{equation}\label{EQ:222}
\dot{\rho}_{2} +3H(\rho_{2} + p_{2}) = - Q,    
\end{equation}
where $\rho_{1}$, $p_{1}$ are the energy density and pressure of the DE and $\rho_{2}$, $p_{2}$ are the energy density and pressure of CS sectors. $Q$ represents the strength of interaction, which may assume arbitrary forms. There are no strict constraints on the sign of $Q$ and depending on its sign, energy may flow from one sector of fluid to the other. In this case, $Q > 0$ corresponds to an energy transfer from the cosmic string sector to the dark energy sector, and $Q < 0$ corresponds to an energy transfer from the DE sector to CS. From Eqs. (\ref{EQ:221}) and (\ref{EQ:222}) it demonstrates that the individual fluids violate the conservation equation while the total energy density of the cosmic  fluids however satisfy the usual form of conservation equation which is \cite{Paul_2015},
\begin{equation}\label{EQ:223}
    \dot{\rho}_{1} + 3H(1 + \omega_{1}^{eff})\rho_{1} = 0
\end{equation}
\begin{equation}\label{EQ:224}
    \dot{\rho}_{2} + 3H(1 + \omega_{2}^{eff})\rho_{3} = 0
\end{equation}
where $\omega_{1}^{eff}$ and $\omega_{3}^{eff}$ are the effective EoS parameters defined as,
\begin{equation}\label{EQ:225}
    \omega_{1}^{eff} = \omega_{1} - \frac{Q}{3H\rho_{1}},\;\; \omega_{2}^{eff} = \omega_{2} + \frac{Q}{3H\rho_{2}},
\end{equation}
where it is evident that the effective EoS parameter now depends on the interaction strength.
Different functional forms of interactions were taken up in the literature. There are no strict rules to assume a particular form of interaction. Some phenomenological choices are made initially, which are then verified using astronomical observations. Several authors have considered different forms of $Q$ such as $Q \propto \rho_{1}$ \cite{PhysRevD.85.043007}, $Q \propto \dot{\rho}_{1}$ \cite{V_liviita_2008}, $Q \propto \rho_{2}$ \cite{10.1111/j.1365-2966.2009.16115.x, di_valentino_can_2017-1}. Cosmological models obtained using several of these interactions are found consistent with the observational results \cite{PhysRevD.96.123508, PhysRevD.97.043529}. Thus, any new interaction form must be constrained using observations to construct a stable cosmological model. In this paper, we consider a linear form of interaction given by,
\begin{equation}\label{EQ:226}
    Q = 3H\eta\rho_{1},
\end{equation} 
where $\eta$ is a coupling parameter that denotes the interaction strength.

Using Eqs. (\ref{EQ:211}), (\ref{EQ:223}) and (\ref{EQ:224}), the total energy density for the cosmic fluid was obtained, which yields
\begin{equation}\label{EQ:227}
\rho = \rho_{10}a^{-3(1 +\omega_{1}^{eff})} + \rho_{20}a^{-3(1 +\omega_{2}^{eff})} + \rho_{30}a^{-3(1 + A)},
\end{equation}
where $\rho_{10} = \frac{B^2}{(1 + A)^2}$, $\rho_{20} = \frac{2BK}{(1 + A)^2}$ and $\rho_{30} = \frac{K^2}{(1 + A)^{2}}$, and the effective EoS parameters are,
\begin{equation}\label{EQ:228}
    \omega_{1}^{eff} = -1 - \eta, \;\; \omega_{2}^{eff} = \omega_{2} + \eta\alpha,
\end{equation}
where $\alpha = \frac{\rho_{1}}{\rho_{2}}$ is a positive quantity and $\omega_{2} = - \frac{1}{3}$. In this paper, we consider $\omega_{2}^{eff} = 0$ for getting a matter-dominated universe. Hence, the value of $\eta$ must be $ > 0$, corresponding to an increasing DE density. For $\eta > 0$, the effective EoS of DE deviates from the non-interacting case and lies in the phantom region.

Finally, the expression for the pressure in an interacting universe can be determined which are
\[
p = - \frac{(1 + \eta)B^2}{(1 + A)^2}\frac{1}{a^{3(1 +\omega_{1}^{eff})}} + \frac{2BK(\omega_{2} + \eta\alpha)}{(1 + A)^2}\frac{1}{a^{3(1 +\omega_{2}^{eff})}}
\] 
\begin{equation}\label{EQ:229}
    + \frac{AK^2}{(1 + A)^2}\frac{1}{a^{3(1 + A)}}.
\end{equation}
\section{Dynamical analysis of the model}
\label{sec:dynamic}
The dynamical system analysis is based on differential equations associated with the time derivatives. There is no unique theory for exploring dynamic systems. The evolution rule governing the dynamical system should thus be examined in various ways to determine its different features. Thus, instead of solving the nonlinear differential equations which are highly nonlinear, we adopt a technique to represent the dynamical equations to analyze the stability of the system. The stability can be examined using various methods, including Jacobi stability, Kosambi-Cartan-Chern (KCC) theory, and Lyapunov methods. In the paper, we shall use the Jacobi stability analysis to perform the dynamical system analysis of the background equations of EU model with non-interacting and interacting cases. 
\subsection{Non-interacting case}
For the study of the EU model with the non-interacting case, we consider below $2$ dynamical variables, $x$ and $y$, which transform the field equations in terms of the dynamical variables as,
\begin{equation}\label{311}
    x = \frac{2BK}{(1 + A)^{2}3H^{2}}\frac{1}{a^{\frac{3}{2}(1 + A)}},
\end{equation}
\begin{equation}\label{312}
    y = \frac{K^{2}}{(1 + A)^{2}3H^{2}}\frac{1}{a^{3(1 + A)}}.
\end{equation}
Using the above dynamical variables, the field Eq. (\ref{EQ:211}) reduces to,
\begin{equation}\label{313}
    \Omega_{DE} = 1 - x - y,
\end{equation}
where $\Omega_{DE} = \frac{B^{2}}{(1 + A)^{2}}\frac{1}{3H^{2}}$ and from Eq. (\ref{EQ:212}), we get
\begin{equation}\label{314}
    \frac{\dot{H}}{H^{2}} = - \frac{3}{2}(1 + A)\left(\frac{x}{2} + y\right).
\end{equation}

Then we can differentiate $x$ and $y$ with respect to $N = \ln a$ which can be rewritten as 
two differential equations:
\begin{equation}\label{315}
    x' = - \frac{3}{2}(1 + A)x - 2x\frac{\dot{H}}{H^{2}},    
\end{equation}
\begin{equation}\label{316}
    y' = - 3(1 + A)y - 2y\frac{\dot{H}}{H^{2}}.    
\end{equation}
Now, we can redefine deceleration parameter $q$ and the total EoS state parameter $\omega_{total}$ as follows:
\begin{equation}\label{317}
    q = - 1 - \frac{\dot{H}}{H^{2}} = - 1 + \frac{3}{2}(1 + A)\left(\frac{x}{2} + y\right)
\end{equation}
and 
\begin{equation}\label{318}
    \omega_{total} = - 1 - \frac{\dot{H}}{H^{2}} = - 1 + (1 + A)\left(\frac{x}{2} + y\right).
\end{equation}
We use the 2D autonomous system of differential equations (\ref{315}) - (\ref{316}) to explore the different features, determining the critical points and carrying out the local stability of these points. The critical points for the systems are: $P(0,1)$, $Q(1,0)$ and $R(0,0)$. Table \ref{tab2} provides the critical points and the cosmological behavior. The detailed description of each critical point has been narrated below for different values of model parameter $A$:
\begin{table*}
\centering
\caption{Critical Points and the corresponding cosmology with non-interacting fluids for different values $A$}
\begin{tabular}{ |c | c | c | c | c | c | c | c | c | c | }
\hline
 $A$ & critical point & $\Omega_{DE}$ & $\Omega_{2}$ & $\Omega_{3}$  & Phase of Universe & $\omega_{total}$ & $q$ & Eigenvalues & Stability\\
\hline
 & $(0,1)$ & $0$ & $0$ & $1$ & Radiation dominated & $\frac{1}{3}$ & $1$ & $\{4,2\}$ & Unstable node \\ 
 $\frac{1}{3}$ & $(1,0)$ & $0$ & $1$ & $0$ & CS dominated & $- \frac{1}{3}$ & $0$ & \{$-2,2\}$ & saddle\\  
 & $(0,0)$ & $1$ & $0$ & $0$ & DE dominated & $-1$ & $-1$ & $\{-4,-2\}$ &  Stable node \\
 \hline
 & $(0,1)$ & $0$ & $0$ & $1$ & CS dominated & $-\frac{1}{3}$ & $0$ & $\{2,1\}$ &  Unstable node \\ 
 $- \frac{1}{3}$ & $(1,0)$ & $0$ & $1$ & $0$ & Domain Wall dominated & $-\frac{2}{3}$ & $- \frac{1}{2}$ & $\{-1,1\}$ & saddle\\  
 & $(0,0)$ & $1$ & $0$ & $0$ & DE dominated & $-1$ & $-1$ & $\{-2,-1\}$ &  Stable node \\
 \hline
 & $(0,1)$ & $0$ & $0$ & $1$ & stiff matter dominated & $1$ & $2$ & $\{6,3\}$&  Unstable node \\ 
 $1$ & $(1,0)$ & $0$ & $1$ & $0$ & dust dominated & $0$ & $\frac{1}{2}$ & $\{-3,3\}$& saddle\\  
 & $(0,0)$ & 1 & $0$ & $0$ & DE dominated & $-1$ & $-1$ & $\{-6,3\}$&  Stable node \\
\hline
 & $(0,1)$ & $0$ & $0$ & $1$ & Dust dominated & $0$& $\frac{1}{2}$ & $\{3, \frac{3}{2}\}$ & Unstable node \\ 
 $0$ & $(1,0)$ & $0$ & $1$ & $0$ & Exotic matter dominated & $-\frac{1}{2}$ & $-\frac{1}{4}$ & $\{-\frac{3}{2},\frac{3}{2}\}$ & saddle\\  
 & $(0,0)$ & $1$ & $0$ & $0$ & DE dominated & $-1$ & $-1$ & $\{-3,-\frac{3}{2}\}$ & Stable node \\
\hline
\end{tabular}
\label{tab2}
\end{table*}

\begin{figure}
    \centering
    \includegraphics[width=7cm]{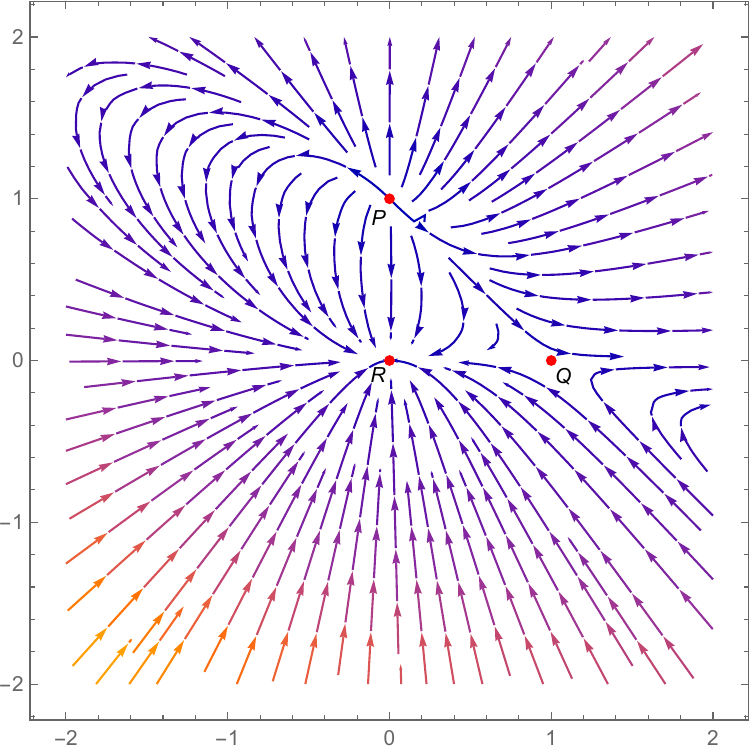}
    \caption{The figure shows 2D phase portrait of the autonomous system for the EU with non-interacting fluids}
    \label{phasespace}
\end{figure}

\begin{itemize}
    \item {\bf Critical point $P(0, 1)$:} For the case $A = \frac{1}{3}$, this point corresponds to the density parameters are, $\Omega_{DE} = 0$, $\Omega_{cs} = 0$ and $\Omega_{r} = 1$. This implies that the critical point becomes absolutely radiation-dominated. The EoS parameter and deceleration parameter are $\omega_{total} = \frac{1}{3}$ and $q = 1$ respectively. This behaviour of the critical point leads to the decelerating phase of the Universe. The corresponding eigenvalues of the Jacobian matrix: $\{4, 2\}$, i.e., they are all positive hence the point is an unstable node. The point $P(0,1)$ also describes the CS-dominated, stiff matter-dominated and dust-dominated universe for $A = -\frac{1}{3}$, $A = 1$ and $A = 0$, respectively. The solution corresponds to an unstable node.  
    \item {\bf Critical point $Q(1, 0)$:} We obtain   a CS dominated universe with $\Omega_{cs} = 1$ for $A = \frac{1}{3}$ at this point. The EoS parameter and deceleration parameter are $\omega_{total} = - \frac{1}{3}$ and $q = 0$ respectively. This behaviour of the critical point leads to the critical phase of the Universe. At the point $Q$, the eigenvalues of the Jacobian matrix: $\{- 2, 2\}$,  they are opposite in sign it corresponds to   a saddle. This point $Q$ is also a saddle point for $A = -\frac{1}{3}$, $A = 1$ and $A = 0$.
    \item {\bf Critical point $R(0, 0)$:} At the point, it corresponds to solution of the DE-dominated phase of the universe with $\Omega_{DE} = 1$. For $A = \frac{1}{3}$, the EoS parameter and deceleration parameters are $\omega_{eff} = - 1$ and $q = - 1$, respectively. Therefore, this critical point leads to the accelerated phase of expansion of the Universe. The eigenvalues of the Jacobian matrix are $\{- 4, - 2\}$, i.e., they are all negative in sign, giving the point $R$ is a stable node.
\end{itemize}

\begin{figure*}
\centering
\begin{subfigure}{0.4\textwidth}
    \includegraphics[width=\textwidth]{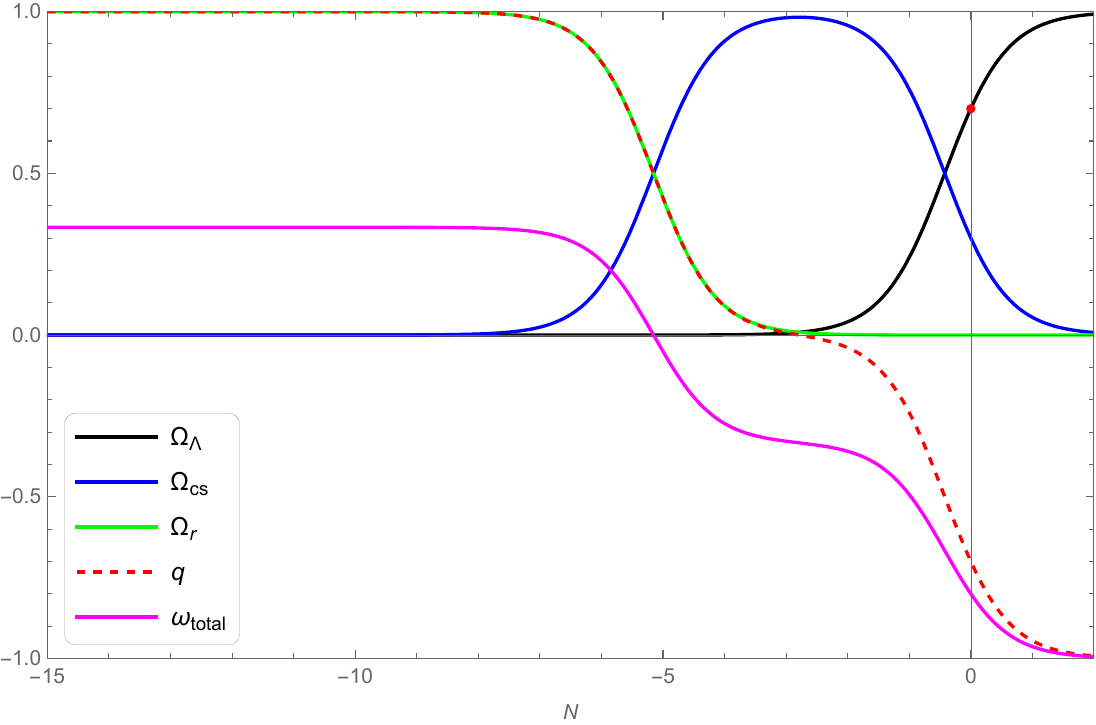}
    \caption{$A = \frac{1}{3}$}
    \label{fig:first}
\end{subfigure}
\hfill
\begin{subfigure}{0.4\textwidth}
    \includegraphics[width=\textwidth]{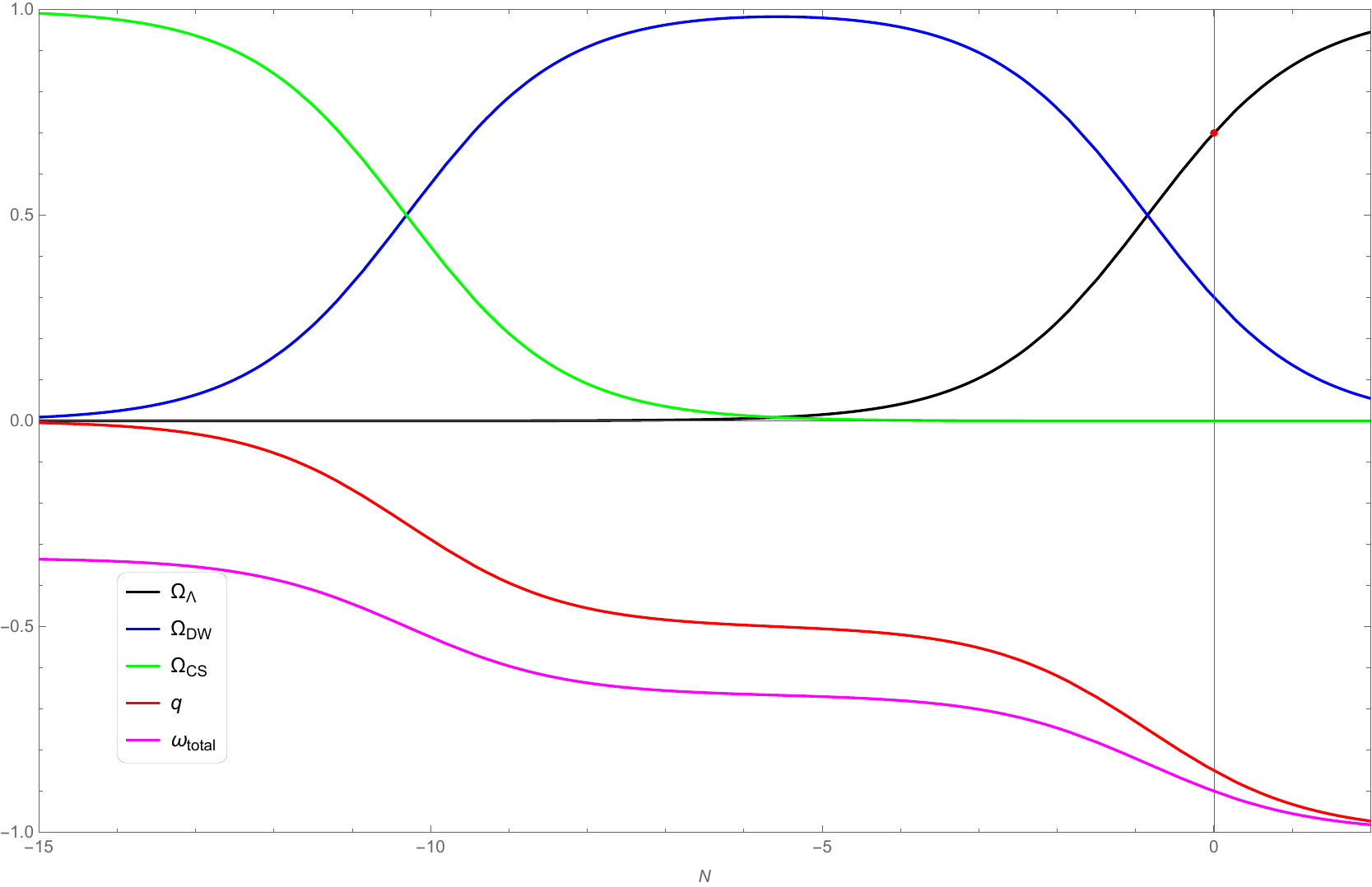}
    \caption{$A = - \frac{1}{3}$}
    \label{fig:second}
\end{subfigure}
\hfill
\begin{subfigure}{0.4\textwidth}
    \includegraphics[width=\textwidth]{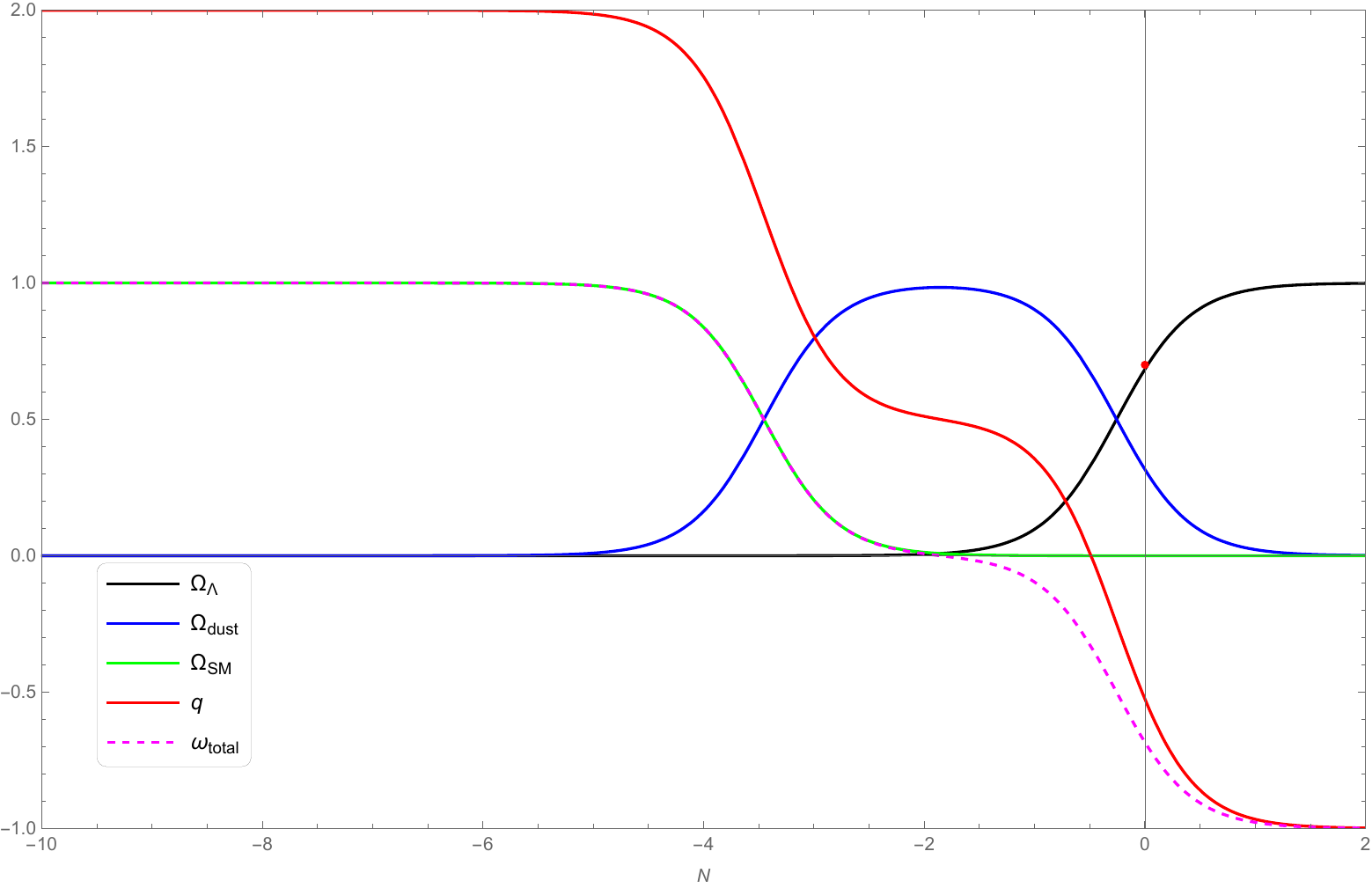}
    \caption{$A = 1$}
    \label{fig:third}
\end{subfigure}
\hfill
\begin{subfigure}{0.4\textwidth}
    \includegraphics[width=\textwidth]{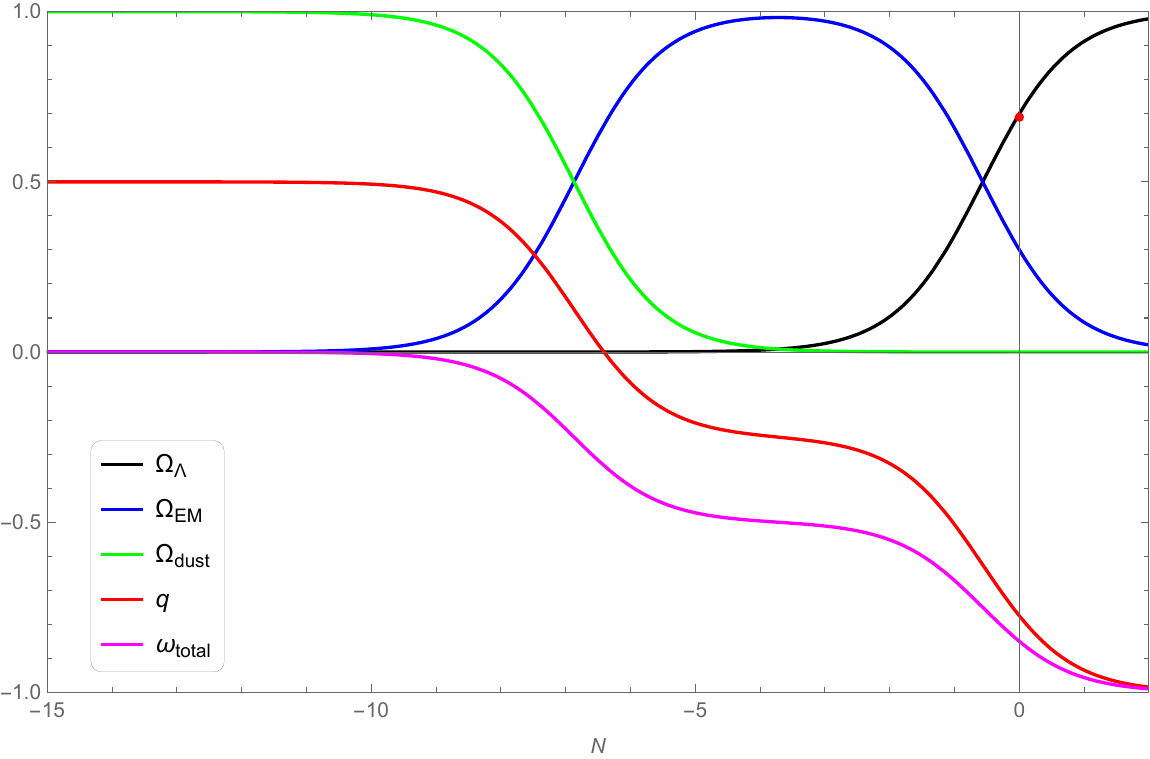}
    \caption{$A = 0$}
    \label{fig:fourth}
\end{subfigure}        
\caption{The evolution of cosmological parameters for the non-interacting case. Panel (a) for $A = \frac{1}{3}$, (b) for $A = - \frac{1}{3}$, (c) for $A = 1$ and (d) for $A = 0$. The red dot is the present value of DE density parameter $(\Omega_{DE} \sim 0.7$).}
\label{dy1}
\end{figure*}

In Fig. \ref{phasespace}, the 2D phase portrait has been given. This shows the trajectory
behaviour, first going from the repeller point $P$ to the saddle point $Q$ and subsequently from $Q$ to the stable point $R$. Further, the evolution plot for cosmological parameters has been given in Fig. \ref{dy1}. From the evolution curve, it is shown that the universe is accelerating at the present epoch. The present value of the effective EoS parameter is $- 1 < \omega_{total} < - \frac{1}{3}$ for all possible values of $A$ in the case of non-interaction. Hence, the Universe shows a quintessence behaviour at the present accelerating phase.

\subsection{Interacting case}
For interacting cases, we consider 3 - parameters, $x$, $y$ and $w$, which are dimensionless and the field equations can be expressed in terms of the dynamical variables as follows:
\begin{equation}\label{321}
    x = \frac{B^{2}}{(1 + A)^{2}}\frac{1}{3H^{2}}a^{-3(1 + \omega_{1}^{eff})},
\end{equation}
\begin{equation}\label{322}
    y = \frac{2BK}{(1 + A)^{2}}\frac{1}{3H^{2}}a^{-3(1 + \omega_{2}^{eff})}, 
\end{equation}
\begin{equation}\label{323}
    w = \frac{K^{2}}{(1 + A)^{2}}\frac{1}{3H^{2}}a^{-3(1 + A)}.
\end{equation}

Therefore, Eq. (\ref{EQ:227}) reduces to,
\begin{equation}\label{324}
    x + y + w = 1,
\end{equation}
and subsequently, from Eq. (\ref{EQ:229}), we get
\begin{equation}\label{325}
    \frac{\dot{H}}{H^{2}} = - \frac{3}{2}\left(1 + \omega_{1}^{eff}x + \omega_{2}^{eff}y + A w\right).
\end{equation}

As we are concerned here with the asymptotic behavior of evolution, we take derivatives of $x$, $y$ and $w$ with respect to the number of e-folding $N = \ln a$. The following system of ordinary differential equations is obtained in terms of dynamical variables:
\begin{equation}\label{326}
    x' = - 3(1 + \omega_{1}^{eff})x - 2x\frac{\dot{H}}{H^{2}},    
\end{equation}
\begin{equation}\label{327}
    y' = - 3(1 + \omega_{2}^{eff})y - 2y\frac{\dot{H}}{H^{2}},    
\end{equation}
\begin{equation}\label{328}
    w' = - 3(1 + A)w - 2w\frac{\dot{H}}{H^{2}}.    
\end{equation}
The evolution of the dynamical variables $x$, $y$ and $w$ corresponding to the variation of the cosmological parameters $\Omega_{DE}$, $\Omega_{m}$ and $\Omega_{r}$, respectively. Finally, we can redefine the deceleration parameter $q$ and the total equation of state parameter $\omega_{total}$ as
\begin{equation}\label{329}
    q = - 1 - \frac{\dot{H}}{H^{2}} = - 1 + \frac{3}{2}\left(1 + \omega_{1}^{eff}x + \omega_{2}^{eff}y + A w\right)
\end{equation}
and 
\begin{equation}\label{329a}
    \omega_{total} = - 1 - \frac{2}{3}\frac{\dot{H}}{H^{2}} = - 1 + \left(1 + \omega_{1}^{eff}x + \omega_{2}^{eff}y + A w\right).
\end{equation}

Similar to the non-interacting case, we shall determine the critical points from 3D autonomous system and carry out the local stability of the EU with interacting fluids. We determine the critical points for the autonomous systems (\ref{326}-\ref{328}) with $P_{1}(0,0,1)$, $P_{2}(0,1,0)$ and $P_{3}(1,0,0)$. Table \ref{tab3} provides these critical points and the cosmological behavior at these points. The detailed description of each critical point is narrated below.
\begin{table*}
\centering
\caption{Critical Points and the corresponding cosmology for interaction with model parameter $A = \frac{1}{3}$, $\omega_{1}^{eff} = - 1.2$ and $\omega_{2}^{eff} = 0$}
\begin{tabular}{ |c | c | c | c | c | c |c| c| }
\hline
 Name & critical point & $\Omega_{DE}$ & $\Omega_{m}$ & $\Omega_{r}$  & Phase of Universe & Eigenvalues & Stability\\
\hline
$P_{1}$ & (0,0,1) & 0 & 0 & 1 & Radiation dominated &[4.6, 1, 1] & Unstable node \\ 
 $P_{2}$ & (0,1,0) & 0 & 1 & 0 & matter dominated & [3.6, -1, 0] & saddle\\  
$P_{3}$ & (1,0,0) & 1 & 0 & 0 & DE dominated & [-4.6, -3.6, -3.6] & Stable node \\
\hline
\end{tabular}
\label{tab3}
\end{table*}

\begin{figure*}
\centering
\begin{subfigure}{0.4\textwidth}
    \includegraphics[width=\textwidth]{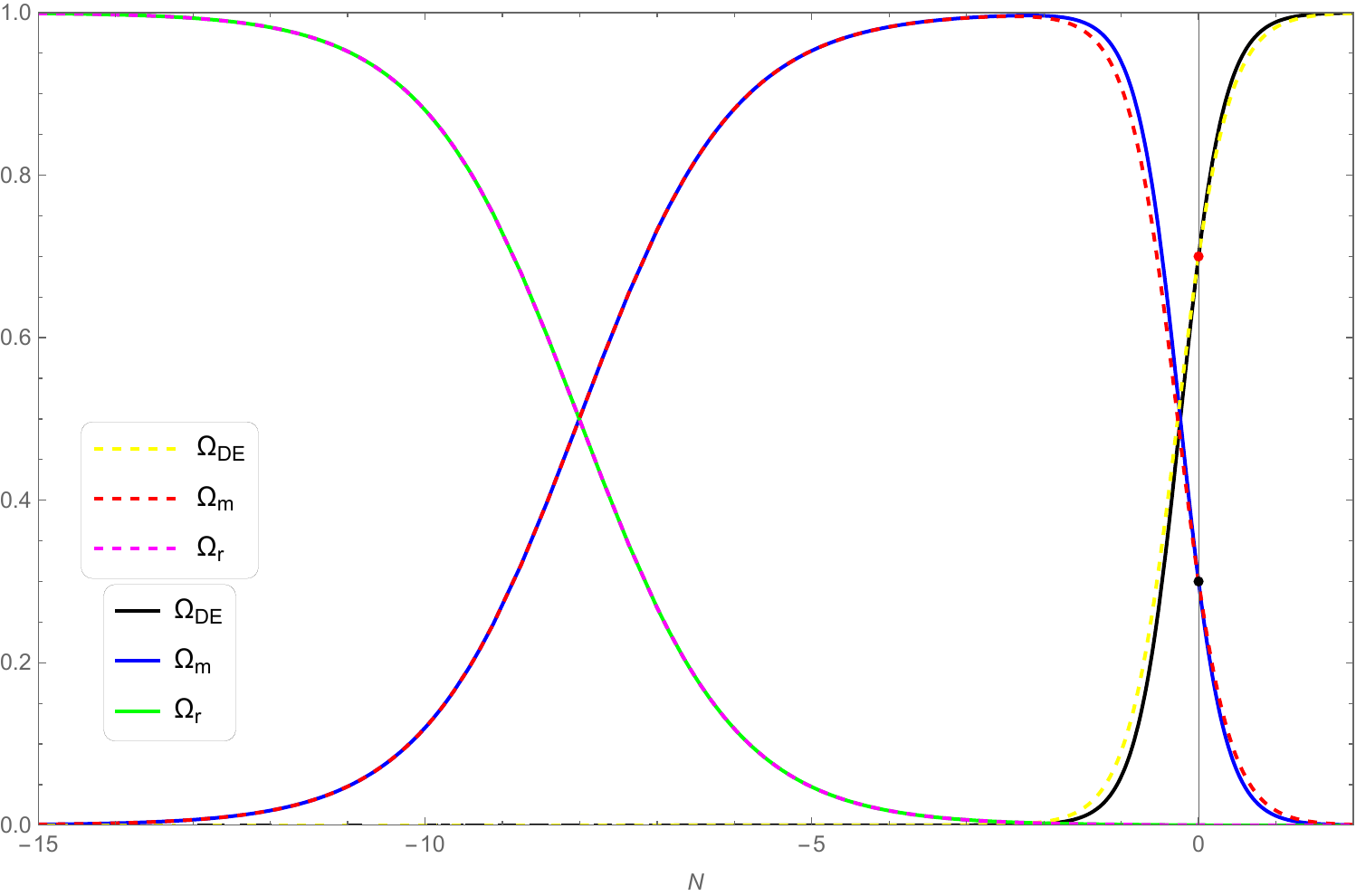}
    \caption{}
    \label{fig:first}
\end{subfigure}
\hfill
\begin{subfigure}{0.4\textwidth}
    \includegraphics[width=\textwidth]{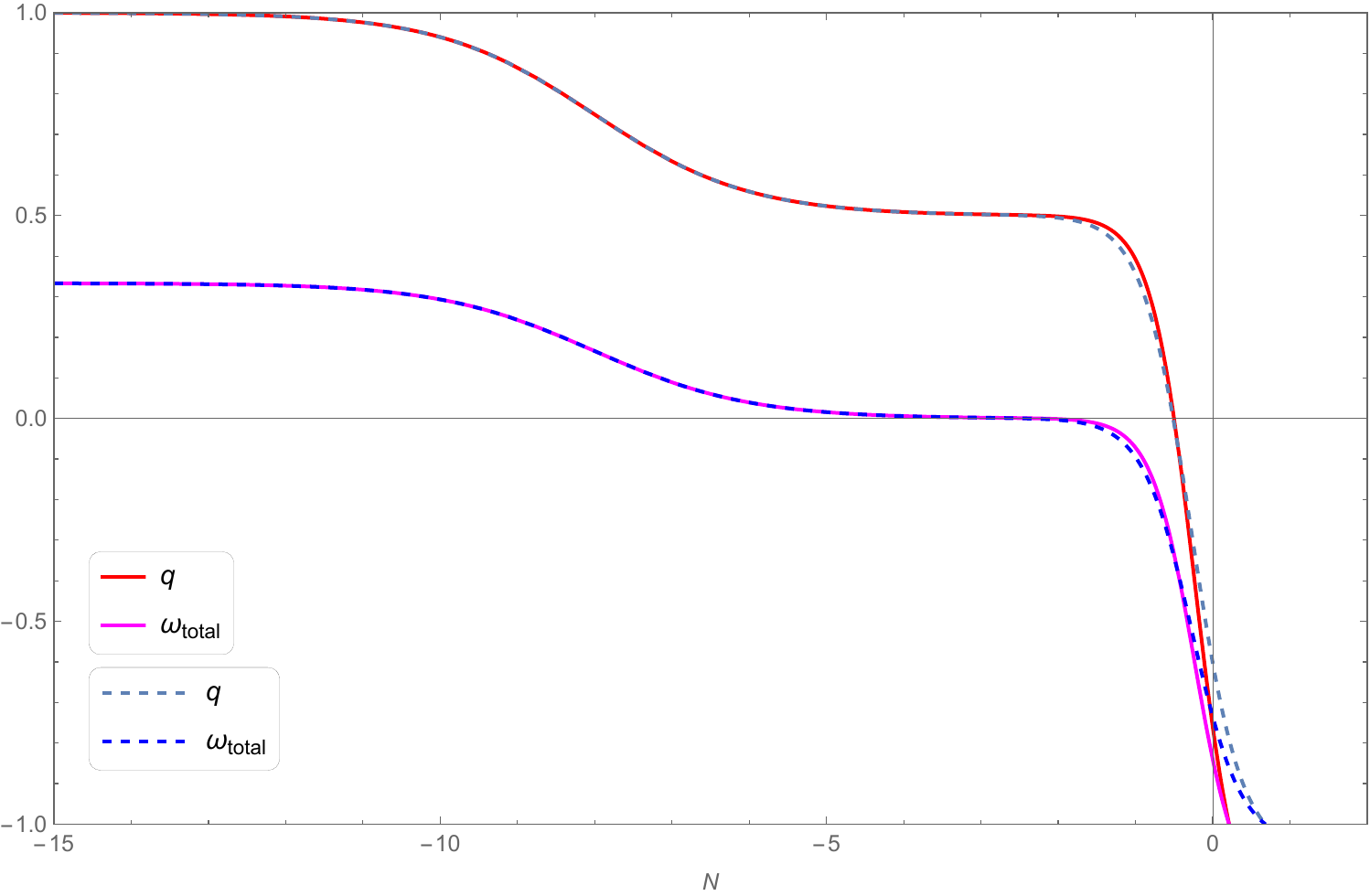}
    \caption{}
    \label{fig:second}
\end{subfigure}
\caption{The figure shows the evolution of cosmological parameters with interaction. In panels (a) and (b), the dashed and solid curves correspond to the interaction strength $\eta = 0.05$ and $\eta = 0.2$, respectively. In panel (a), the red dot is the present value of DE density parameter $(\Omega_{DE} \sim 0.7$).}
\label{dy_int}
\end{figure*}

\begin{itemize}
    \item {\bf Critical point $P_{1}(0, 0, 1)$:} For the point the corresponding density parameters are, $\Omega_{DE} = 0$, $\Omega_{m} = 0$ and $\Omega_{r} = 1$. This implies that the critical point becomes absolutely radiation-dominated. The EoS parameter and deceleration parameter are $\omega_{total} = \frac{1}{3}$ and $q = 1$ respectively. This behavior of the critical point leads to the decelerating phase of the Universe. At the point, eigenvalues of the Jacobian matrix: $\{4.6, 1, 1\}$, i.e., they are all positive hence the point is an unstable node.
    \item {\bf Critical point $P_{2}(0, 1, 0)$:} Corresponding density parameters are, $\Omega_{DE} = 0$, $\Omega_{m} = 1$ and $\Omega_{r} = 0$. This implies that the critical point becomes absolutely CS-dominated. The EoS parameter and deceleration parameter are $\omega_{total} = 0$ and $q = \frac{1}{2}$ respectively. This behaviour of the critical point leads to the critical phase of the Universe. At point $P_{2}$, eigenvalues of the Jacobian matrix: $\{3.6, - 1, 0\}$, i.e., they are opposite in sign hence the point is a saddle.
    \item {\bf Critical point $P_{3}(1, 0, 0)$:} The density parameters are: $\Omega_{DE} = 1$, $\Omega_{m} = 0$ and $\Omega_{r} = 0$, which corresponds to the fact that this critical point is absolutely DE-dominated phase. The EoS parameter and deceleration parameter are $\omega_{total} = - 1 - \eta$ and $q = - 1 - \frac{3}{2}\eta$ respectively. Therefore, this critical point leads to the accelerated DE-dominated phase of the Universe for $\eta > 0$. The eigenvalues of the Jacobian matrix at the point C: $\{- 4.6, - 3.6, -3.6\}$, i.e., they are all negative in sign hence the point is a stable node.
\end{itemize}
\begin{figure*}
\centering
\begin{subfigure}{0.4\textwidth}
    \includegraphics[width=\textwidth]{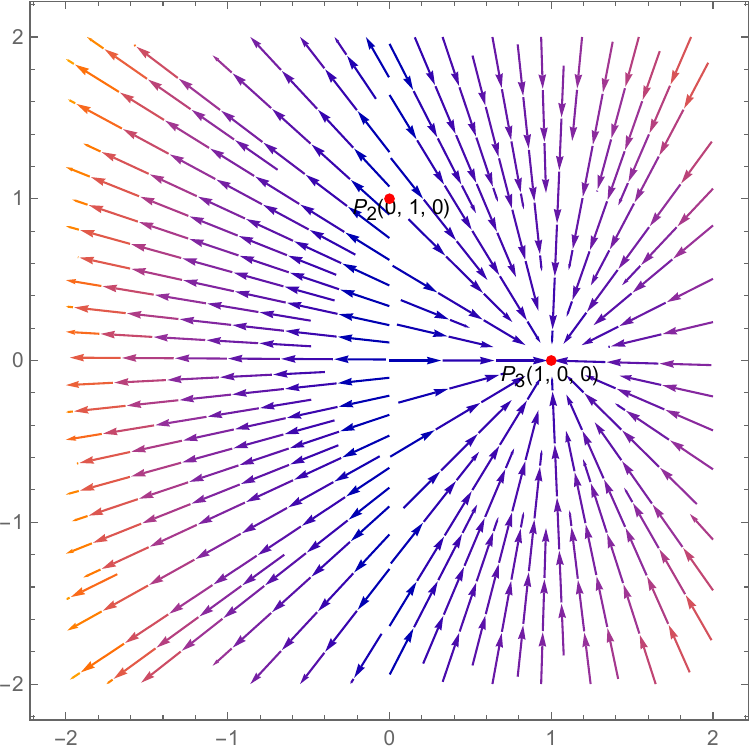}
    \caption{}
    \label{fig:first}
\end{subfigure}
\hfill
\begin{subfigure}{0.4\textwidth}
    \includegraphics[width=\textwidth]{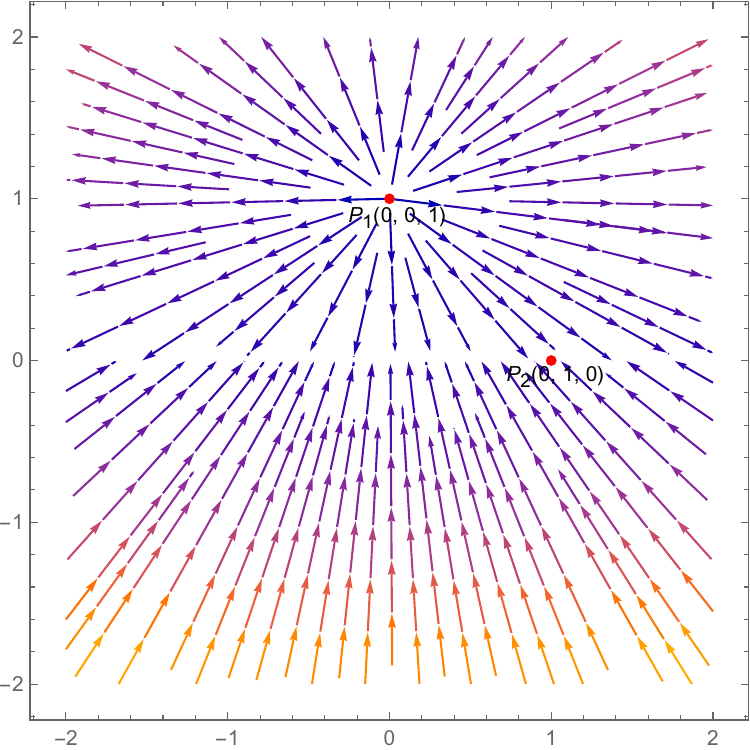}
    \caption{}
    \label{fig:second}
\end{subfigure}
\hfill
\begin{subfigure}{0.4\textwidth}
    \includegraphics[width=\textwidth]{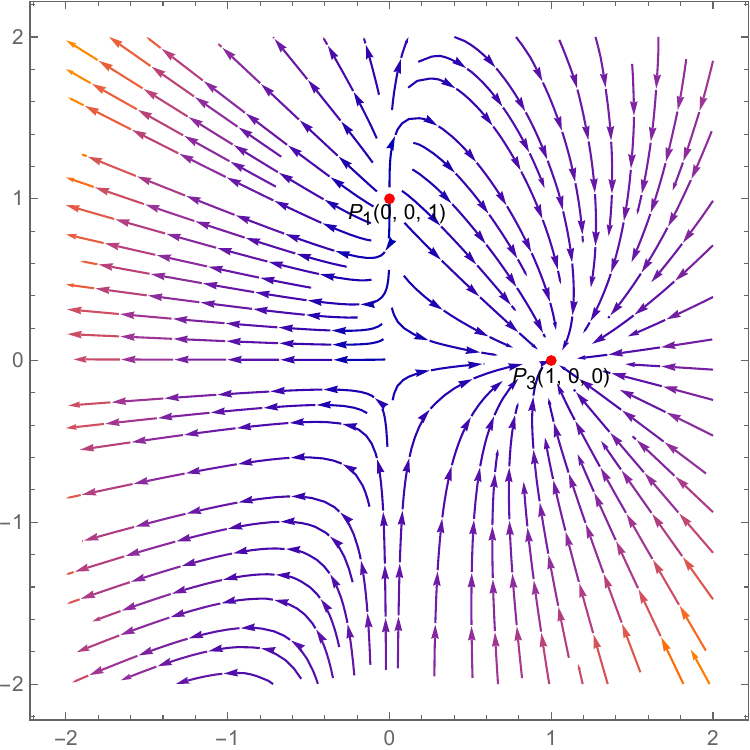}
    \caption{}
    \label{fig:third}
\end{subfigure}        
\caption{Shows the 2D phase portrait of the autonomous system for interacting fluids. The phase planes for the values of $\omega_{1}^{eff} = -1.2$, $\omega_{2}^{eff} = 0$, $A = \frac{1}{3}$ and $\eta = 0.2$. Panel (a) the portrait on the $x-y$ plane, (b) the portrait on the $y-z$ plane and (c) the portrait on the $x-z$ plane.}
\label{phase_inter}
\end{figure*}
In Fig. \ref{phase_inter}, the 2D phase portrait is drawn to understand the stability of the points. The panels display the trajectories for critical points, where $P_{1}$ acts as the attractor, absorbing all incoming trajectories, and $P_{3}$ acts as the repeller, repelling all incoming trajectories. As $P_{2}$ is a saddle, it repels trajectories that originate from itself and absorbs trajectories from $P_{1}$. Hence, the critical point $P_{1}$ is an unstable node, whereas $P_{3}$ is a stable node. The $P_{2}$ is an unstable saddle point. Further, the evolution of cosmological parameters has been given in Fig. \ref{dy_int} for two different interaction strengths. It is demonstrated that for the strengths $0.05$ and $0.2$, the current values of the effective EoS parameter are $-0.75$ and $-0.84$, respectively.
\section{Conclusions}
\label{sec:conclusions}
In the present work, we present a general framework of the emergent universe scenario with a non-linear equation of state in the general theory of relativity. The scale factor is finite at infinite past for this emergent universe model. In this case, the Emergent universe is effectively composed of three types of fluids to compare the importance, we consider both the interacting and non-interacting fluids to understand the observed universe. For $A = \frac{1}{3}$, the dark energy, cosmic string and radiation are the cosmic fluids components of the universe without interaction. When interaction sets in, depending on the strength of interaction, the cosmic fluids are found to transform into three other types, say,  dark energy, pressureless matter, and radiation, which can describe the present observed universe.

Here we adopt a dynamical autonomous system analysis for the EU with and without interaction among the cosmic fluids. In the case of non-interaction, we first transform the background evolution equations into an autonomous system considering two dynamical parameters. For this case, the critical points are found at $P(0,1)$, $Q(1,0)$ and $R(0,0)$. For $A = \frac{1}{3}$, the point $P$ is an unstable node with all eigenvalues for the Jacobian matrix positive where  $Q$ represents an unstable saddle point with all eigenvalues opposite in sign for the Jacobian matrix. The point $R$ with all the eigenvalues for the Jacobian matrix are negative in the sign, giving the stable node behaviour. In table \ref{tab2}, we study in detail the characteristics of each critical point for all possible values of the model parameter for a given  $A$. The 2D phase portrait has been shown in Fig. \ref{phasespace}, from which the nature of the critical points can be ascertained clearly. It is found that a transition from an unstable node to a stable node followed by a saddle point is permitted. The evolutionary behaviour of the cosmological parameter in the case of non-interacting fluid is shown in Fig. \ref{dy1}. For each case, the evolutionary curve of the dark energy density parameter crosses the present observed value. We find a stable point in the future when the universe will be determined completely by dark energy.

Similar to the non-interacting case, we get more interesting results if the interaction is considered.  In the interacting case in the EU, the critical points corresponding to the autonomous systems are $P_{1}(0,0,1)$, $P_{2}(0,1,0)$ and $P_{3}(1,0,0)$. The characteristic of each critical point is discussed in the table \ref{tab3}. The point $P_{1}$ with all eigenvalues positive gives a radiation-dominated unstable node. A mater-dominated saddle point $P_{2}$ is also found in this model, which is followed by the transition to a stable dark energy-dominated universe. In Fig. \ref{dy_int}, one can see that the total EoS parameter is close to the observational value for interaction strength $\eta = 0.2$. The variation of the deceleration parameter shows an accelerating phase of the universe, which is followed by a transition from the deceleration phase.
\section*{Acknowledgement}
BCR, AC, and BCP would like to thank the IUCAA Centre for Astronomy Research and Development (ICARD), NBU, for extending research facilities. BCR and BCP gratefully acknowledge IUCAA, Pune, for its invaluable support and provision of research facilities, which significantly contributed to this study. BCR also thanks the Ministry of Social Justice and Empowerment, Govt. of India, and the University Grants Commission (UGC), India, for their financial support. BCP would like to thank SERB DST Govt. of India for a project grant(F. No. CRG/2021/000183).
\section*{Data Availability}
There is no new data associated with this article.
\vspace{1cm}

\bibliographystyle{apsrev4-1}
\bibliography{Refs}

\end{document}